\documentclass[lettersize,journal]{IEEEtran}

\usepackage{amsmath,amsfonts,amssymb}
\usepackage{array}
\usepackage{booktabs}
\usepackage{textcomp}
\usepackage{stfloats}
\usepackage{url}
\usepackage{graphicx}
\usepackage{placeins}
\usepackage{cite}
\providecommand{\tightlist}{\setlength{\itemsep}{0pt}\setlength{\parskip}{0pt}}

\begin{document}

\title{Recovery Rates Are Not Comparable Across Transcription Factors:
Chance Correction for Attribution Evaluation}

\author{Hyunkyung~Han
        and~Min~Jung~Kim
\thanks{H. Han is with the Department of Integrative Medicine, Yonsei
University College of Medicine, Seoul 06273, Republic of Korea
(e-mail: sthan1@yonsei.ac.kr).}%
\thanks{M. J. Kim is with the Department of Integrative Medicine, Yonsei
University College of Medicine, Seoul 06273, and with the Department of
Radiology, Research Institute of Radiologic Science, Yonsei University
College of Medicine, Seoul 03722, Republic of Korea
(e-mail: MINES@yuhs.ac). ORCID: H. Han, 0009-0006-2672-384X;
M. J. Kim, 0000-0003-4949-1237.}%
\thanks{Preprint. This work has not been peer reviewed.}}

\markboth{Preprint,~2026}%
{Han \MakeLowercase{\textit{et al.}}: Chance Correction for Attribution Evaluation}

\maketitle

\begin{abstract}
Attribution methods for genomic sequence models are commonly evaluated
by how much of a known motif they recover, or by how a prediction
degrades as evidence is deleted. Neither score is interpretable without
the value it would take by chance, and neither is routinely reported
against one. We show that this omission is not a matter of precision but
of validity. The uniform chance level for contiguous motif overlap is
\(L/(N-L+1)\); across 268 transcription factors in UniBind it ranges
from 0.0118 to 0.0427, a 3.6-fold spread determined by motif length and
window size alone. For two factors the bootstrap intervals of the chance
levels themselves do not overlap, so their raw recovery rates are not
comparable quantities. Correcting for this dissolves a published
three-way classification of five factors: a factor reported as a
resolution failure attains the second-highest corrected value, ahead of
one of the two positive controls, and two reported as complete failures
fall at or below chance.

We further show that perturbation-based evaluation can fail its own
precondition: for one factor a fully masked input still scores above the
decision boundary, and the curve is not monotone in the number of masked
positions, so the area under it is not a measure of faithfulness. We
provide chance levels in closed form, a chance-corrected score, and two
screens that run before any attribution is computed.
\end{abstract}


\IEEEpeerreviewmaketitle

\begin{figure*}[!t]
\centering
\includegraphics[width=\textwidth]{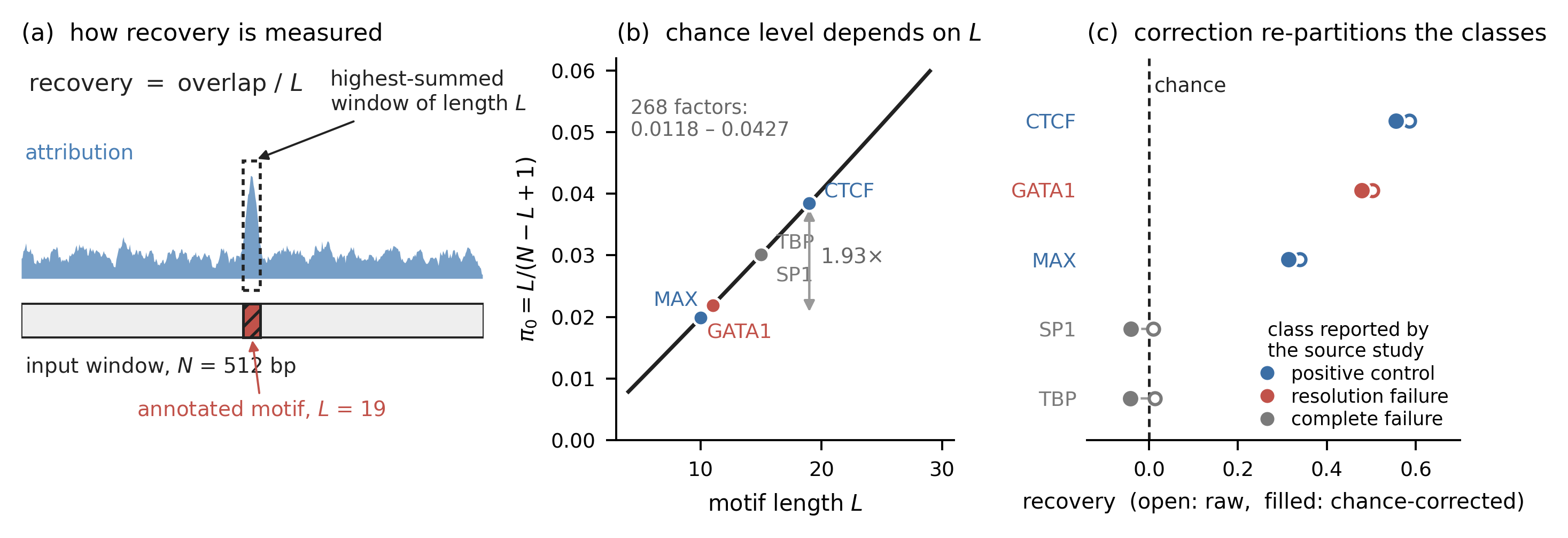}
\caption{Recovery rates are measured against a baseline that is not
constant. \textbf{(a)}~The score is the overlap between the annotated
motif and the highest-summed attribution window of the same length,
divided by that length. \textbf{(b)}~The value such a window attains by
chance is $\pi_0=L/(N-L+1)$, which varies with motif length; across 268
factors it spans a 3.6-fold range, and the two factors used as positive
controls differ by a factor of 1.93. \textbf{(c)}~Correcting each factor
against its own chance level moves the reported three-way classification:
the factor reported as a resolution failure attains the second-highest
corrected value, and the two reported as complete failures fall at or
below zero. Open markers are raw recovery, filled markers are
chance-corrected.}
\label{fig:overview}
\end{figure*}


\subsection{1. Introduction}

Attribution methods are increasingly used to argue that a genomic
sequence model has learned biology rather than an artefact. The argument
usually takes one of two forms: the attribution overlaps a known binding
motif, or deleting the highest-attributed positions degrades the
prediction. Both are reported as raw numbers.

A raw overlap of 0.06 is not a claim until one knows what overlap an
uninformative attribution would achieve. That value is not zero, and it
is not constant. It depends on the length of the motif and on the width
of the window the model was given, both of which vary between datasets
and between factors within a dataset. The same is true of inter-method
agreement scores, which are sometimes read as though 0.5 were a natural
midpoint.

Our concern is not that the numbers are imprecise. It is that they are
not comparable. Across 268 transcription factors the uniform chance
level spans a 3.6-fold range, and the spread is driven entirely by motif
geometry. For the five factors we measure in detail, the bootstrap
intervals of the chance levels for CTCF and MAX do not overlap.
Correcting for the chance level then re-partitions a published set of
conclusions rather than refining it.

A second and more basic failure affects perturbation curves. Such a
curve presupposes that removing all of the evidence yields a prediction
at or below the decision boundary; otherwise the curve is bounded below
by a quantity unrelated to the attribution. We find that this
precondition does not always hold, and that it is checkable in a single
forward pass.

None of the individual ingredients is novel. The correction we use has
the form of the adjusted Rand index, where the same problem --- an
overlap statistic that is not zero under independence --- was addressed
by subtracting the expectation and rescaling (Hubert and Arabie, 1985).
The observation that the choice of null model determines the conclusion
was formulated in data mining (Hanhijärvi et al., 2009; Lijffijt et al.,
2014) and carried into genomics (De et al., 2014; Ferkingstad et al.,
2015; Kanduri et al., 2019). Chikina and Troyanskaya (2012) computed an
exact combinatorial null for interval proximity and argued from it that
raw proximity statistics are not comparable across datasets; Salvatore
et al.~(2020) made the corresponding argument for similarity measures in
genomic colocalisation. Dinucleotide-preserving shuffles date to
Altschul and Erickson (1985) and have been generally available since
uShuffle (Jiang et al., 2008). What has not been done is to apply any of
this to the evaluation of attributions, where recovery rates are still
reported against an implicit baseline of zero. We do so, quantify what
it changes, and release the implementation.

\begin{center}\rule{0.5\linewidth}{0.5pt}\end{center}

\subsection{2. Chance levels}

\begin{figure*}[!t]
\centering
\IfFileExists{F1_null_distribution.png}%
  {\includegraphics[width=0.98\textwidth]{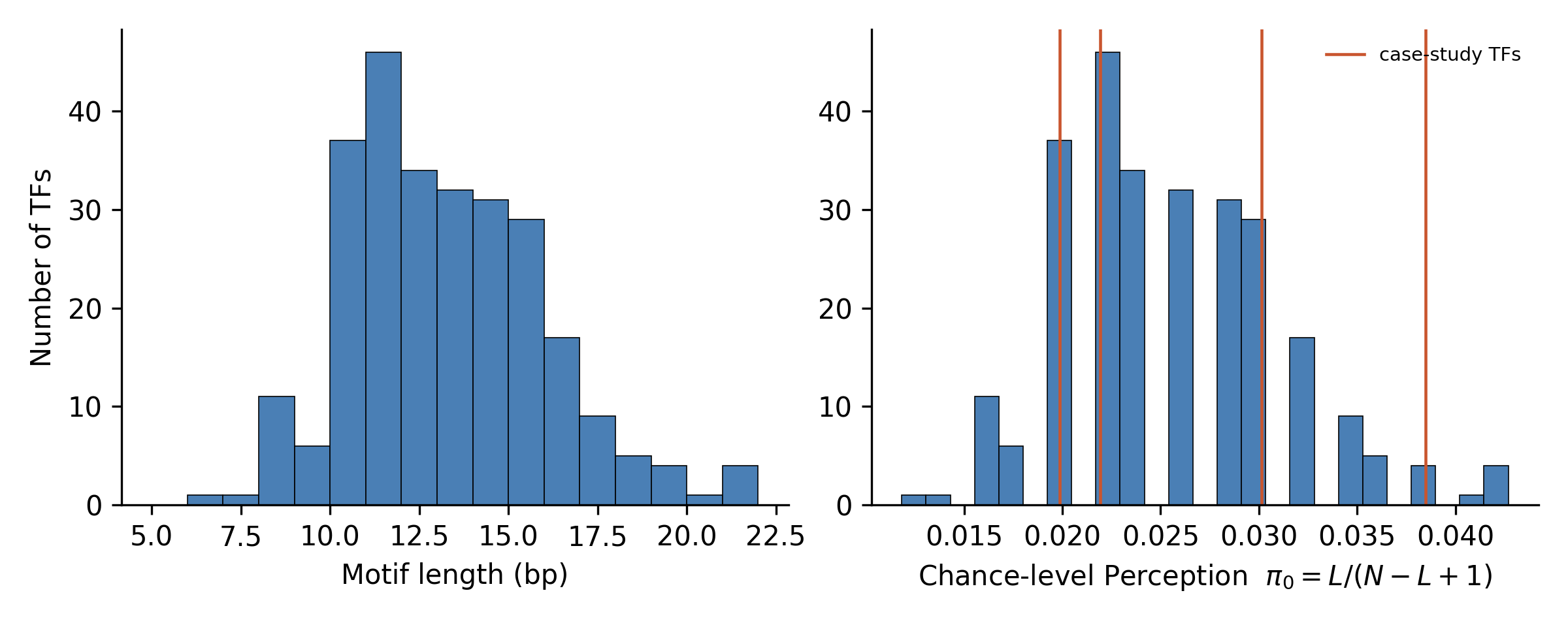}}%
  {\fbox{\rule{0.90\textwidth}{0pt}\rule{0pt}{45mm}}}
\caption{Null distributions for contiguous motif overlap. \textbf{Left:} annotated motif lengths across 268 factors. \textbf{Right:} the resulting chance levels $\pi_0=L/(N-L+1)$; the five detail factors are marked. SP1 and TBP share a motif length and therefore coincide.}
\label{fig:null}
\end{figure*}

\subsubsection{2.1 Contiguous overlap}

For a motif of length \(L\) placed uniformly at random in a window of
length \(N\), the expected contiguous overlap with a fixed region of
length \(L\) is

\[\pi_0 = \frac{L}{N - L + 1}.\]

We verified this against simulation to within 2.6\% for
\(L = 6 \ldots 25\) at \(N = 512\) over 200,000 draws.

\subsubsection{2.2 Positional null}

Where motifs concentrate near the window centre --- which happens
whenever windows are cropped around peak summits --- the uniform null is
too loose. We therefore also compute a positional null
\(\pi_{\mathrm{pos}}\) by simulation from the empirical distribution of
motif start positions. For CTCF at \(L = 19\), \(N = 512\), this gives
0.0697 against a uniform null of 0.0385. Position alone, with no
sequence information, accounts for a 1.59--2.02-fold enrichment over the
uniform null across the five factors.

\subsubsection{2.3 Inter-method
agreement}

The Jaccard index between the top-\(k\) features selected by two
attribution methods has chance value 0.020 for \(k = 20\) in a window of
512. Both sets have identical fixed cardinality by construction, so the
size-driven bias documented by Salvatore et al.~(2020) does not arise
here.

\subsubsection{2.4 Adjusted score}

\[\mathrm{APS} = \frac{s - \pi}{1 - \pi}\]

zero at chance, one at perfect recovery, negative below chance. This is
the correction form of the adjusted Rand index (Hubert and Arabie, 1985)
applied to the overlap statistic.

A negative value should not be read as an attribution that avoids the
motif. At the magnitudes we report (\(-0.042\) and \(-0.041\)), the
recovery is indistinguishable from what an attribution ignoring the
sequence entirely would achieve; the sign carries no information at that
distance from zero. We treat such values as uninformative rather than as
anti-correlated, in the same way we treat a correlation computed on a
near-constant prediction.

\begin{center}\rule{0.5\linewidth}{0.5pt}\end{center}

\subsection{3. Two screens}

\subsubsection{3.1 Composition screen}

Given a classifier and its positive and negative sets, we compare
performance on the originals with performance on dinucleotide-preserving
shuffles of the positives. If shuffled sequences score at or above the
originals, the task does not require sequence order, and no attribution
method can report motif use that the model did not make.

\subsubsection{3.2 Masking baseline}

We evaluate the model on a fully masked input. If that prediction sits
above the decision boundary, the deletion curve is bounded below by a
value unrelated to the attribution.

\begin{center}\rule{0.5\linewidth}{0.5pt}\end{center}

\subsection{4. Setup}

\begin{itemize}
\tightlist
\item
  \textbf{Data.} UniBind, 268 transcription factors. Positive windows of
  512 bp centred on binding sites; negatives are GC-matched genomic
  windows.
\item
  \textbf{Models.} HyenaDNA and DNABERT-2, fine-tuned per factor.
\item
  \textbf{Attribution.} Integrated gradients.
\item
  \textbf{Detail set.} Five factors measured in full: CTCF, GATA1, MAX,
  SP1, TBP.
\end{itemize}

Each model was fine-tuned per factor for three epochs with AdamW
(learning rate \(3\times10^{-5}\), weight decay 0.01, one-cycle
schedule), batch size 32, and a maximum sequence length of 512.

\subsubsection{GC matching}

GC-matched genomic backgrounds are a standard recommendation, but the
recommendation does not specify a matching rule, and the rule matters.
Our implementation matches the GC content histogram of the positive set
in bins of width 0.02, drawing candidate windows until the histogram is
filled. Six of the 268 factors (CLOCK, ELF5, HINFP, NR5A1, OCT4, SOX9)
are excluded before matching because their positive sets fall below the
minimum size the screen requires. For the remaining 262 the histogram
rule filled the background to full size in every case. We report the
rule because it is not conventionally reported, and a matching procedure
that silently returns a smaller or differently distributed background
will change the recovery rates computed against it.

\begin{center}\rule{0.5\linewidth}{0.5pt}\end{center}

\subsection{5. Results}

\subsubsection{5.1 Chance levels are not comparable across
factors}

Across 268 factors the uniform chance level ranges from 0.0118 to
0.0427, a 3.6-fold spread; motif lengths run from 6 to 21 with a median
of 12. Restricted to the five factors examined in the source study the
spread is 1.85-fold, so the problem is not an artefact of the wider set.
The spread is a function of motif length and window size and carries no
information about any model.

In the five-factor benchmark the annotated motif length is constant
within each factor, and the window is 512 bp throughout, so the uniform
chance level is a deterministic quantity with no sampling uncertainty
attached to it.

\begin{table}[!t]
\caption{Chance levels for the five detail factors. \(L\) is the annotated motif length, constant across all sequences of a given factor.}
\label{tab:t1}
\centering
\setlength{\tabcolsep}{4pt}
\small
\begin{tabular}{lcccc}
\toprule
TF & sequences & \(L\) & uniform null & centre-biased null \\
\midrule
MAX & 31,557 & 10 & 0.0199 & 0.0488 \\
GATA1 & 10,404 & 11 & 0.0219 & 0.0455 \\
SP1 & 1,808 & 15 & 0.0301 & 0.0688 \\
TBP & 2,313 & 15 & 0.0301 & 0.0658 \\
CTCF & 42,736 & 19 & 0.0385 & 0.0888 \\
\bottomrule
\end{tabular}
\end{table}

The chance level for CTCF is 1.93 times that for MAX. This is not an
estimate with an interval around it: given the motif length and the
window, it is the value an uninformative attribution attains in
expectation. A reader comparing raw recovery rates for these two factors
is comparing quantities measured against baselines that differ by a
factor of two, and the difference is a property of motif geometry rather
than of anything either model learned. SP1 and TBP, which share a motif
length, share a chance level exactly.

We verified these values against simulation with uniform random
importance: the closed form falls inside the bootstrap interval of the
simulated null for all five factors.

\begin{figure*}[!t]
\centering
\IfFileExists{F5_reclassification.png}%
  {\includegraphics[width=0.82\textwidth]{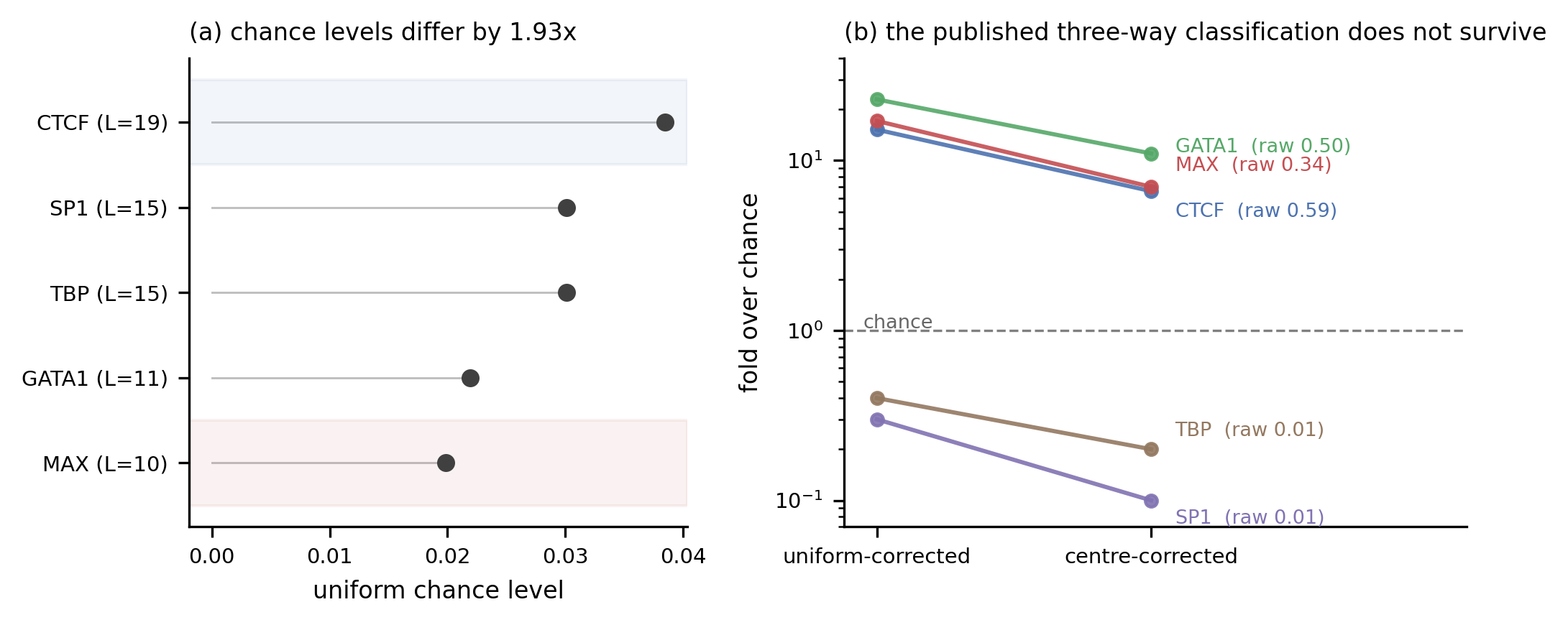}}%
  {\fbox{\rule{0.90\textwidth}{0pt}\rule{0pt}{45mm}}}
\caption{Re-partition. \textbf{(a)} Uniform chance levels for the five factors, which differ by 1.93-fold between the two positive controls. \textbf{(b)} Fold over chance under the uniform and the centre-biased null. GATA1, reported as a resolution failure, is indistinguishable from the positive controls; SP1 and TBP fall below one under either correction.}
\label{fig:repart}
\end{figure*}

\subsubsection{5.2 A published classification does not survive
correction}

The source study sorts five factors into three classes and uses them as
stress tests: CTCF and MAX as positive controls binding long,
high-signal motifs; SP1 and TBP as compositional-bias tests, selected as
a GC-rich and an AT-rich case; and GATA1 as a resolution test, on the
grounds that it binds a short and often degenerate motif. It reports
that recovery is good for the positive controls, degraded for GATA1, and
close to absent for SP1 and TBP.

We measure the same five factors on our own fine-tuned models and
correct each against its own chance level. The partition does not
survive (Fig.~\ref{fig:repart}).

\begin{table}[!t]
\caption{Motif recovery on our models, before and after correction. The null is the positional one, estimated from the empirical distribution of motif starts in the same data.}
\label{tab:t2}
\centering
\setlength{\tabcolsep}{4pt}
\small
\begin{tabular}{lcccc}
\toprule
TF & recovery & \(\pi_{\mathrm{pos}}\) & adjusted & reported class \\
\midrule
CTCF & 0.586 & 0.0697 & \textbf{+0.555} & positive control \\
GATA1 & 0.502 & 0.0443 & \textbf{+0.479} & resolution failure \\
MAX & 0.339 & 0.0364 & \textbf{+0.314} & positive control \\
TBP & 0.013 & 0.0531 & \textbf{$-$0.042} & complete failure \\
SP1 & 0.009 & 0.0478 & \textbf{$-$0.041} & complete failure \\
\bottomrule
\end{tabular}
\end{table}

GATA1, the resolution test, is the second-best recovered of the five and
is closer to CTCF than to either of the factors it was grouped with in
failure. SP1 and TBP sit at or just below zero, meaning their recovery
is not degraded but uninformative: an attribution that ignored the
sequence entirely would score the same. What remains is not three
classes but two --- three factors recovered above chance, two not
recovered at all --- and the boundary does not fall where the stress
tests placed it.

Two of the three class definitions also fail on their own terms. The
compositional-bias pair is supposed to contrast a GC-rich case with an
AT-rich one, but the GC content of the two positive sets is 0.580 and
0.582 (Section 5.3). And the resolution test is justified by a motif of
roughly six base pairs, while every GATA1 motif in the annotation file
the study distributes is eleven base pairs long.

The same re-partition follows from the study's own reported values.
Reading its Figure 4a and dividing by the corresponding chance levels
gives 15.3, 15.4, 13.3, 1.4 and 1.5 under the uniform null for CTCF,
MAX, GATA1, SP1 and TBP, and 6.2, 6.2, 6.6, 0.7 and 0.8 under the
positional one: GATA1 indistinguishable from the positive controls, SP1
and TBP at or below chance. We report this as a consistency check rather
than as evidence, because those values are read from a published figure
and the code that produced them is not part of the study's release. For
the same reason we cannot determine how sequences carrying more than one
annotated motif --- about five percent for MAX --- were handled. Where
the two accounts differ in ordering among the three recovered factors,
we make no claim: the difference is within what we can resolve.

\begin{figure}[!t]
\centering
\IfFileExists{F2_gap_vs_aps.png}%
  {\includegraphics[width=\columnwidth]{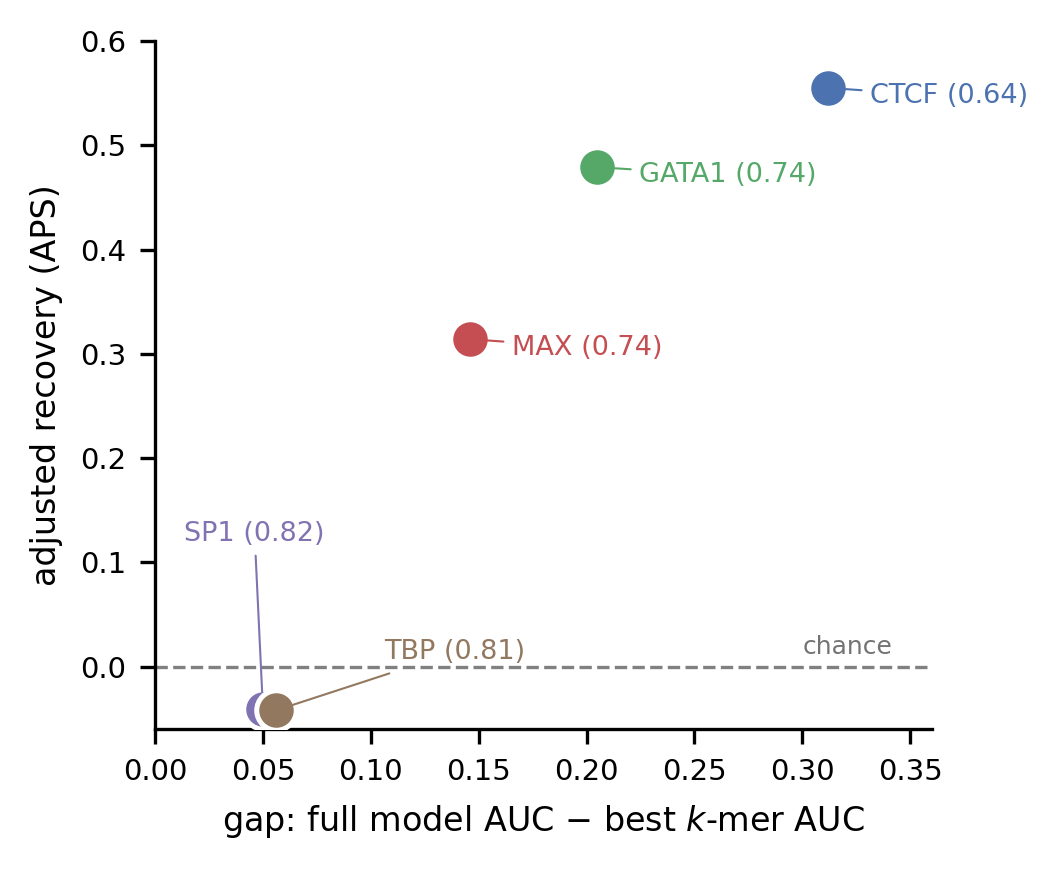}}%
  {\fbox{\rule{0.93\columnwidth}{0pt}\rule{0pt}{52mm}}}
\caption{Composition gap versus adjusted recovery. The horizontal axis is the difference between the full model's AUC and the best $k$-mer probe; values in parentheses are the best $k$-mer AUC. Where the gap is small the task is close to solvable by composition alone, and the adjusted score sits at chance.}
\label{fig:gap}
\end{figure}

\begin{figure*}[!t]
\centering
\IfFileExists{F3_bimodal.png}%
  {\includegraphics[width=0.98\textwidth]{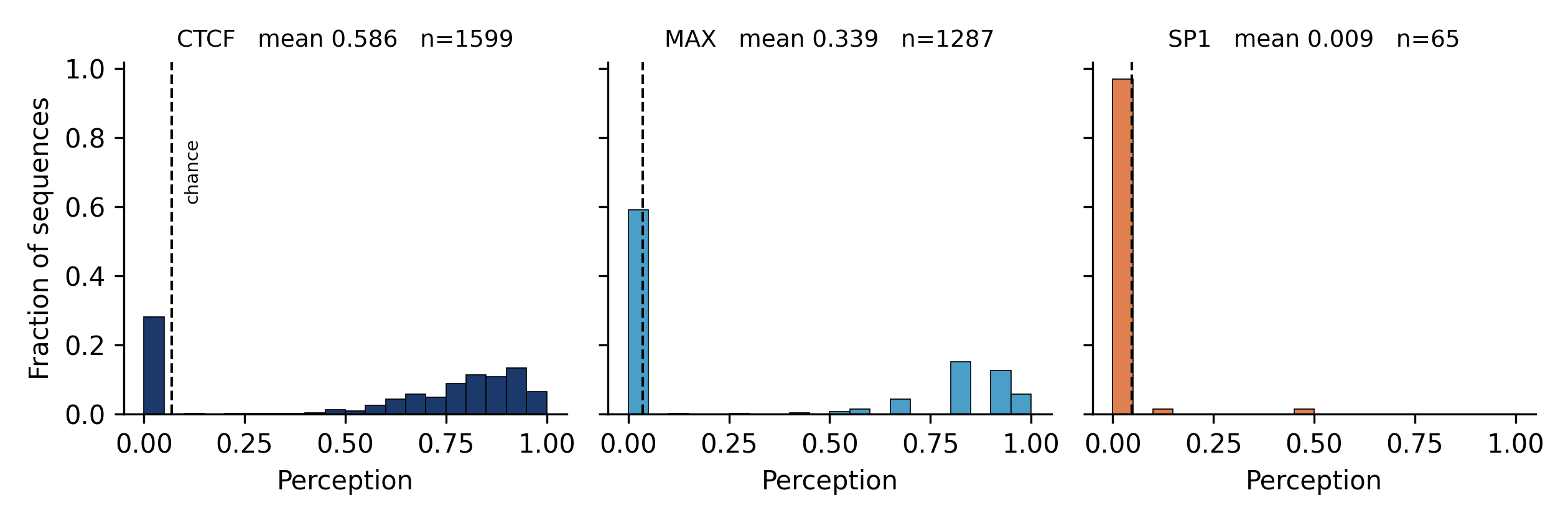}}%
  {\fbox{\rule{0.90\textwidth}{0pt}\rule{0pt}{40mm}}}
\caption{Per-sequence recovery is bimodal. For each factor the score is either near zero or near one rather than concentrated at its mean, so the mean should be read alongside the fraction of sequences above 0.5.}
\label{fig:bimodal}
\end{figure*}

\begin{figure*}[!t]
\centering
\IfFileExists{F4_ablation.png}%
  {\includegraphics[width=0.98\textwidth]{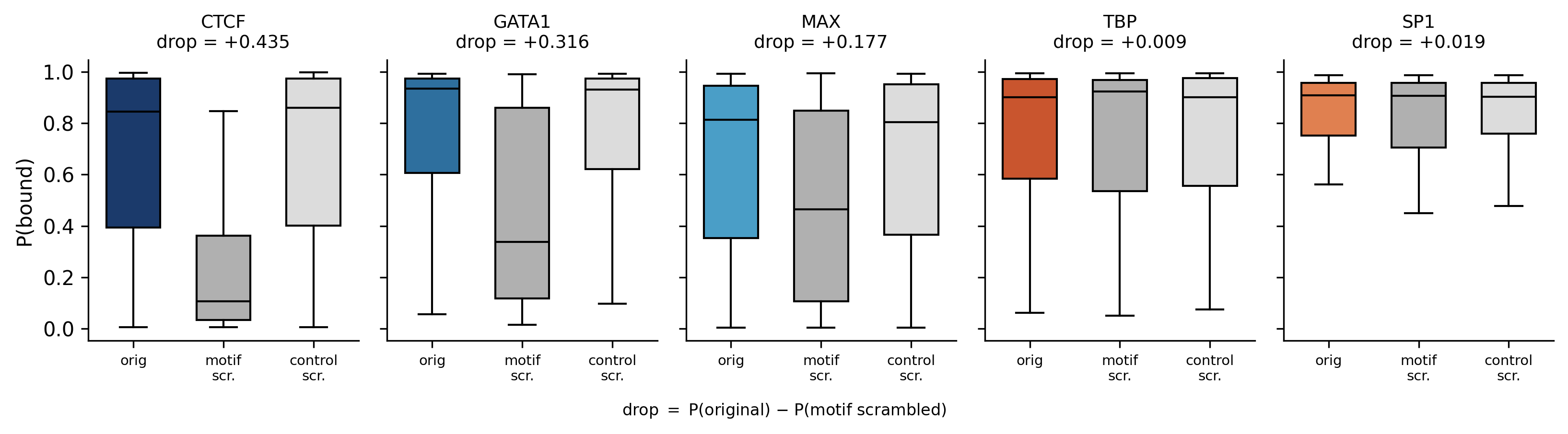}}%
  {\fbox{\rule{0.90\textwidth}{0pt}\rule{0pt}{40mm}}}
\caption{Motif ablation. Predicted probability on the original sequence, with the motif replaced by composition-matched bases, and with an unrelated region of the same length replaced. The motif-specific drop is the difference between the second and third boxes.}
\label{fig:ablate}
\end{figure*}

\subsubsection{5.3 The models are not using the motif where the task
does not require
it}

\begin{table}[!t]
\caption{Composition screen and adjusted scores, measured on our own fine-tuned models. The first column is the gap between the classifier's score on dinucleotide-shuffled positives and on the negatives; the second is the gap between the full model and the best $k$-mer probe. These are not the values in Table~\ref{tab:t2}, which are corrections of the source study's reported numbers; the two are computed from different models and should not be compared entry by entry.}
\label{tab:t3}
\centering
\setlength{\tabcolsep}{3.5pt}
\footnotesize
\begin{tabular}{lccccc}
\toprule
      & shuffled     & $k$-mer &          &                       &     \\
TF    & $-$ negative & gap     & recovery & $\pi_{\mathrm{pos}}$ & APS \\
\midrule
CTCF  & $+0.238$ & 0.319 & 0.586 & 0.0697 & $\mathbf{+0.555}$ \\
GATA1 & $+0.420$ & 0.240 & 0.502 & 0.0443 & $+0.479$ \\
MAX   & $+0.694$ & 0.202 & 0.339 & 0.0364 & $+0.314$ \\
TBP   & $+0.724$ & 0.095 & 0.013 & 0.0531 & $\mathbf{-0.042}$ \\
SP1   & $+0.635$ & 0.094 & 0.009 & 0.0478 & $\mathbf{-0.041}$ \\
\bottomrule
\end{tabular}
\end{table}

The ordering is legible (Fig.~\ref{fig:gap}). The two factors with the
smallest gap between the full model and the best $k$-mer probe --- SP1
at 0.094 and TBP at 0.095, against 0.319 for CTCF --- are the two whose
adjusted score is at or just below zero. Where the task is closest to
being solvable by composition alone, the attribution is not failing;
there is nothing for it to find. The first column of
Table~\ref{tab:t3}, which compares the classifier on shuffled positives
against the negatives, orders the factors similarly but not identically:
TBP and MAX have the largest values there.

The argmax position makes the same point directly: for CTCF the
highest-attributed position falls at 230 against a motif at 236, whereas
for SP1 it falls at 5 against a motif at 244.

The contrast the source study builds between these factors does not hold
in the data. SP1 is treated as GC-rich and TBP as AT-rich, but the GC
content of their positive windows is 0.580 and 0.582 ---
indistinguishable. A TATA box is AT-rich, but a 512 bp window centred on
a TATA-containing promoter is a CpG island. The two factors the study
contrasts are, at the level of composition, the same case.

Nor do the negatives explain the pattern. The GC difference between
positive and negative sets is 0.006 to 0.021, and a classifier using GC
content alone reaches an AUC of 0.522 to 0.569. Whatever separates the
classes, it is not the background composition.

A blunter version of the same test makes the point without any shuffle.
Replacing the input with a uniformly random sequence should destroy the
signal. For TBP it does not: a random sequence scores 0.983 against
0.745 for the true positives, and for SP1 0.946 against 0.917. These are
the two factors whose adjusted recovery is at or below zero.

\begin{table}[!t]
\caption{Classification AUC on held-out test data.}
\label{tab:t4}
\centering
\setlength{\tabcolsep}{4pt}
\small
\begin{tabular}{lcc}
\toprule
TF & DNABERT-2 & HyenaDNA \\
\midrule
CTCF & 0.9728 & 0.9574 \\
GATA1 & 0.9757 & 0.9755 \\
MAX & 0.9535 & 0.9407 \\
SP1 & 0.9505 & 0.9156 \\
TBP & 0.9357 & 0.9018 \\
\bottomrule
\end{tabular}
\end{table}

Accuracy is high and nearly uniform across factors. It does not
distinguish the factors whose motifs are recovered from those whose
motifs are not, which is the point: accuracy is not evidence that a
model is using the motif.

\begin{figure*}[!t]
\centering
\IfFileExists{F6_deletion_curve.png}%
  {\includegraphics[width=1.00\textwidth]{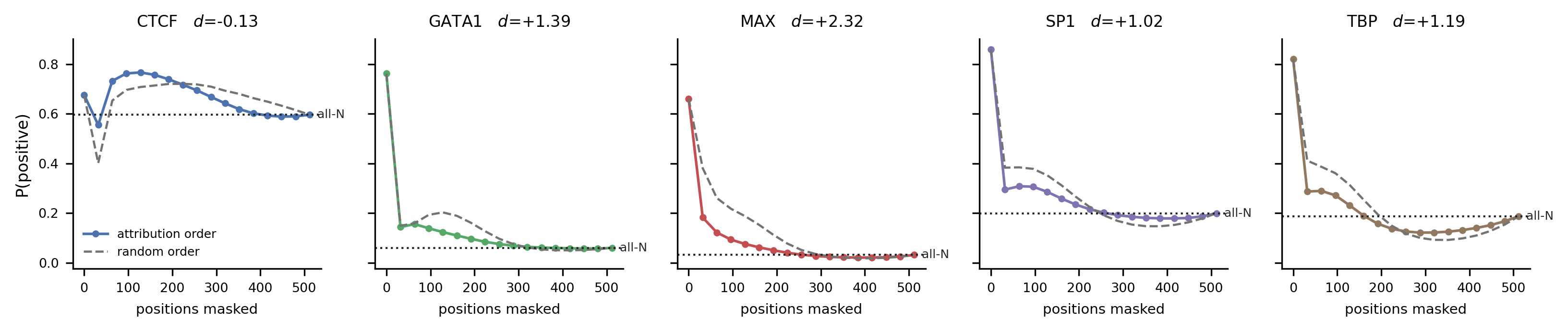}}%
  {\fbox{\rule{0.90\textwidth}{0pt}\rule{0pt}{45mm}}}
\caption{Deletion curves for the five factors, with the all-N convergence value shown as a dotted line. For CTCF the curve rises over most of its extent and converges to 0.596, above the decision boundary.}
\label{fig:delcurve}
\end{figure*}

\subsubsection{5.4 A perturbation curve can fail its own
precondition}

A perturbation curve measures how a prediction falls as evidence is
removed. It presupposes that removing all of the evidence produces a
prediction at or below the decision boundary. For CTCF this is not the
case: a fully masked input scores 0.596, above the threshold. The
deletion curve therefore converges not to zero but to that value,
leaving an AUC dynamic range of 0.165. The area under such a curve is
not a measure of faithfulness.

The condition is not universal, which is what makes it worth checking
rather than assuming: across the five factors the fully masked score
ranges from 0.032 (MAX) to 0.596 (CTCF). A deletion curve computed for
MAX is interpretable; the same curve computed for CTCF is not.

\begin{table}[!t]
\caption{Model output on the original input, on a fully masked (all-N)
input, and on a uniformly random sequence. The deletion curves in
Fig.~\ref{fig:delcurve} are computed on a subsample of 200 sequences per
factor and their unmasked means therefore differ from the first column;
for CTCF the curve starts at 0.676 against 0.797 here.}
\label{tab:t5}
\centering
\setlength{\tabcolsep}{4pt}
\small
\begin{tabular}{lccc}
\toprule
TF & original & all-N & all-random \\
\midrule
CTCF & 0.797 & \textbf{0.596} & 0.303 \\
GATA1 & 0.950 & 0.060 & 0.718 \\
MAX & 0.873 & 0.032 & 0.861 \\
TBP & 0.745 & 0.187 & \textbf{0.983} \\
SP1 & 0.917 & 0.200 & \textbf{0.946} \\
\bottomrule
\end{tabular}
\end{table}

The CTCF curve does worse than lack range: it is not monotone, and over
most of its extent it sits \emph{above} the unmasked prediction.
Starting at 0.676 with no masking, it falls to 0.555 at 32 masked
positions, rises to 0.763 at 96 and peaks at 0.767 at 128 --- a quarter
of the sequence removed --- before declining to the all-N value of
0.596. A quantity that increases as evidence is deleted is not measuring
the deletion of evidence. The masking is not subtracting information; it
is substituting one signal for another, and for this factor the
substitute reads as positive.

The random-order curve behaves the same way and, at the first step,
falls further than the attribution-ordered curve (0.401 against 0.555).
Removing the positions an attribution method calls important perturbs
the prediction \emph{less} than removing arbitrary positions, which is
what the negative effect size in Table~\ref{tab:t6} reports.

\subsubsection{5.5 Deletion curves are not comparable across
tokenizers}

The deletion procedure replaces one nucleotide at a time with N. For a
byte-pair tokenizer this changes the number of tokens the model sees.
Averaged over 20 random 512 bp sequences, DNABERT-2 encodes an unmasked
window in 111.5 tokens and 142.3, 186.2, 253.1 and 445.9 tokens after
20, 50, 100 and 300 positions have been masked --- expansions of 1.28,
1.67, 2.27 and 4.00. At full masking the tokenization collapses to
single characters (514.0 tokens), converging on HyenaDNA, which
tokenizes at single-nucleotide resolution and stays at 513 throughout.
The variability is largest in the middle of the range (sd 7.9 at 100
masked positions), so the confound depends on which positions are masked
as well as how many.

The range over which the source study reports separation between
attribution methods, 20 to 50 masked positions, is already a 1.28- to
1.67-fold expansion.

The deletion area therefore measures different quantities in the two
models: for HyenaDNA, the response to removing evidence; for DNABERT-2,
that response confounded with a change in how the input is segmented.
Because the confound does not depend on the order in which positions are
masked, it compresses the differences between attribution methods, which
is consistent with the near-ceiling clustering of the values reported
for SP1 and TBP.

This is not a new observation, and it has already been handled
elsewhere. Kurth et al.~(2026), comparing AttnLRP explanations of
DNABERT-2 against LRP explanations of a CNN, perturb the transformer at
token level and the convolutional model at nucleotide level while
holding the number of perturbed nucleotides equal in each round. That is
the control the comparison requires. It is absent from the audit we
re-analyse, where models with different tokenizers are compared on the
same deletion curve, and we state the corresponding check as a reporting
requirement.

\subsubsection{5.6 Faithfulness measured against the right
baseline}

\begin{table}[!t]
\caption{Deletion-curve effect size (Cohen's \(d\), random deletion versus attribution-ordered deletion), with N-token masking.}
\label{tab:t6}
\centering
\setlength{\tabcolsep}{4pt}
\small
\begin{tabular}{lc}
\toprule
TF & \(d\) \\
\midrule
CTCF & $-$0.13 \\
SP1 & +1.02 \\
TBP & +1.19 \\
GATA1 & +1.39 \\
MAX & +2.32 \\
\bottomrule
\end{tabular}
\end{table}

CTCF is the factor whose masking baseline fails, and it is also the only
factor whose deletion effect is indistinguishable from zero ---
consistent with the curve having no range to move in.

\subsubsection{5.7 Two hypotheses we
rejected}

We tested and rejected two alternative explanations for the pattern
above.

\textbf{Training-data bias.} If recovery were driven by whether a
binding site appeared in training, held-out and matched evaluation sets
would differ. They do not: the largest difference across factors is
0.023.

\textbf{Compositional confounding of the negatives.} If the GC-matched
negatives were poorly matched, a GC-only classifier would separate them.
It does not: GC-only AUC ranges 0.522--0.569.

\begin{center}\rule{0.5\linewidth}{0.5pt}\end{center}

\subsection{6. Discussion}

\subsubsection{6.1 What is new}

None of the individual ingredients is new; the application is. Recovery
rates in attribution evaluation continue to be reported against an
implicit baseline of zero, four decades after the
dinucleotide-preserving shuffle was described and more than a decade
after the non-comparability argument was made for adjacent quantities.

Our contribution is not the null value itself, which is elementary, but
the demonstration that it matters --- and specifically that correcting
for it re-partitions a set of published conclusions rather than nudging
them.

\subsubsection{6.2 The uniform null as an origin, not as a
test}

The uniform null is used here as an origin, not as a test. Ferkingstad
et al.~(2015) show that the choice of null model in genomics orders the
resulting p values, and that a model preserving more of the data's
structure yields larger ones; they also caution that writing down a null
distribution directly, as we do, is where errors are easily made. Their
analysis is what licenses our use and bounds it. In their Section 3 they
argue that the expectation of the test statistic can reasonably be taken
as invariant across the null models in the hierarchy, and that what
changes with preservation level is its variance.

We use only the first half of that statement. The chance level enters
our score as a shift of origin, and a shift of origin is determined by
the expectation. It does not enter as a variance, and we therefore do
not use it to compute p values, confidence intervals, or any claim that
a recovery rate is significantly above chance. Those judgements require
a null model that preserves dinucleotide composition, and we obtain them
from shuffled controls throughout. Our simulation check validates the
expectation under the uniform null and nothing else.

A reader may object that motif occurrences cluster in cis-regulatory
modules and repetitive elements, so that a uniform placement assumption
is naive. This is correct as a description of the genome and is
precisely the objection Ferkingstad et al.~raise against
uniform-position nulls used as tests. It does not apply to the use made
here. The adjusted score is an effect size, not a test statistic;
clustering changes how much a recovery rate varies from sequence to
sequence, but it does not change where the origin sits, and the origin
is all we take from the uniform null.

\subsubsection{6.3 Where the failure is
located}

Koo and Eddy (2019) established that a model can attain high
classification accuracy while its attributions fail to match a known
motif, and located the cause in the architecture: filter size and
pooling determine whether first-layer filters learn whole motifs or
partial ones, so a distributed representation need not align with the
motif that generated the data. The mechanism we describe is upstream of
the architecture. When the classification task is separable by
composition alone, no architecture is required to use the motif, and an
attribution method that faithfully reports what the model used will
report that the motif was not used. The failure is then correctly
located in the task, not in the interpretability method and not in the
network.

\subsubsection{6.4 Relation to the colocalisation
literature}

The correction we apply has an established counterpart. The Forbes
coefficient normalizes observed co-occurrence by its expectation as a
ratio, where we normalize as a difference; Salvatore et al.~(2020)
recommend it over the raw Jaccard index for genomic colocalisation, on
the grounds that an unnormalized overlap is dominated by dataset size.
The argument we make for attribution scores is theirs, moved to a
setting where it has not been made.

Chikina and Troyanskaya (2012) made the same argument for interval
proximity: because the chance of proximity depends on properties of the
reference set that carry no biological meaning, raw proximity statistics
cannot be compared across datasets, and they computed the exact
probability combinatorially rather than by permutation. We take it in a
different direction: they map the statistic to a p value, whereas we
correct the effect size and leave significance to a preservation-matched
null.

The Jaccard index appears in these literatures in unrelated roles.
Salvatore et al.~use it to compare genomic tracks, where its value is
dominated by track size; we use it to measure agreement between the
top-\(k\) features selected by two attribution methods, where both sets
have identical fixed cardinality. The quantities share a name and a
formula and nothing else.

\subsubsection{6.5 The recommendation-to-practice
gap}

The observation that the choice of null model determines the conclusion
is not new, and it did not originate in genomics. It was formulated in
data mining, where imposing a constraint on a randomization is
understood to raise the p value of the data (Hanhijärvi et al., 2009;
Lijffijt et al., 2014). In genomics it was stated plainly by De et
al.~(2014), who showed that different permutation strategies produce
different null distributions for the same enrichment question; it was
formalized by Ferkingstad et al.~(2015) and carried forward by Kanduri
et al.~(2019), who are of the same group and whose recommendations
should be read as one line rather than two.

Shuffles that preserve dinucleotide composition are not new either. The
Euler-tour permutation that generates them dates to Altschul and
Erickson (1985), who showed that it preserves dinucleotide frequencies
exactly rather than in expectation; a general implementation preserving
arbitrary \(k\)-let counts has been available since uShuffle (Jiang et
al., 2008); and dinucleotide-shuffled controls are routine in ChIP-seq
motif discovery (Bailey, 2011).

The gap is not one of awareness, and it is visible within a single
paper. The audit we re-analyse proposes a tiered set of guidelines for
evaluating interpretability in genomics. Its second guideline, G2,
requires that importance scores be compared against a random baseline,
and that shuffled-sequence controls be considered for sequence data. The
motif-overlap scores in its own Section 3.3 are reported against
neither. The recommendation and the practice appear in the same
document.

We raise this as a correction rather than an objection. The diagnosis in
that paper is right, and the guidelines are the ones we would want the
field to adopt. A demonstration case is exactly where they should be
visible.

\subsubsection{6.6 Reporting checklist}

We close with what we would ask of a paper that reports either kind of
score.

\begin{enumerate}
\def\labelenumi{\arabic{enumi}.}
\tightlist
\item
  The prediction on a fully masked input, stated before any perturbation
  curve is shown.
\item
  The chance level and the adjusted value, for every overlap metric
  reported.
\item
  The result of a composition screen: whether the classification task is
  separable without sequence order.
\item
  The perturbation unit, whenever models with different tokenizers are
  compared.
\end{enumerate}

None of these requires an experiment. The first is one forward pass, the
second is a closed-form expression, the third is a standard shuffle, and
the fourth is a sentence.

\subsubsection{6.7 Scope}

A reader may ask whether the failure in Section 5.4 would disappear
under a different perturbation operator --- modifying the attention mask
rather than substituting a token, or marginalising over substitutions
instead of fixing one. It might. Our subject is the deletion curve as it
is currently computed and reported in this literature, and the check we
propose is a check on that practice. Designing a perturbation operator
that keeps the input in distribution is a separate question, and one we
do not take up here.

The same boundary applies to the composition screen. We show that a task
separable by composition alone gives an attribution method nothing to
find; we do not propose how to construct a task that is not so
separable. That is the subject of Section 6.8.

\subsubsection{6.8 Limitations}

The corrected values in Table~\ref{tab:t2} are computed from figures read off a
published plot, and are therefore reported to the precision that
permits. Our detail analysis covers five factors and two architectures;
the 268-factor result concerns chance levels alone and not model
behaviour. The positional null is estimated from the empirical
distribution of motif starts in the same data, and is therefore not
independent of it.

A further limitation concerns run-to-run variation. Each adjusted
recovery score reported in Tables~\ref{tab:t2} and~\ref{tab:t3} comes
from a single fine-tuning run. Table~\ref{tab:t8} gives the score across
independent runs for each factor. For the two factors we ran five times
the spread is 0.175 (CTCF) and 0.220 (GATA1); for the three we ran three
times it is between 0.022 and 0.050. The comparison between those two
groups is itself informative: with three seeds GATA1 appeared to be the
most stable factor, with a spread of 0.009, and two further seeds
returned values 0.17 below the first three, with validation AUC
unchanged to within 0.004. Three runs are therefore not enough to
characterise this variation.

The between-factor differences on which our re-partition rests are 0.61,
which exceeds the largest observed spread by a factor of nearly three,
so that conclusion is unaffected. Comparisons of similar or smaller
magnitude are not safe on a single run, and the bootstrap intervals
computed within one run do not capture this source of variation: for
CTCF the within-run interval has a half-width of 0.019 against a
between-run standard deviation of 0.066.

Finally, the held-out set for SP1 contains 65 annotated positives, an
order of magnitude fewer than for the other factors, and its
per-sequence distribution in Fig.~\ref{fig:bimodal} should be read with
that in mind.

\subsubsection{6.9 What follows}

What we have shown is that recovery rates are determined in part by task
design. The consequence is a benchmark problem: without datasets whose
negatives are controlled so that motif use is necessary, an attribution
evaluation cannot be said to measure the attribution.

\begin{center}\rule{0.5\linewidth}{0.5pt}\end{center}

\subsection{7. Conclusion}

A motif recovery rate is not a claim until the value it would take by
chance is stated alongside it. That value is not zero, it is not
constant across factors, and for the two factors we examine most closely
its confidence intervals do not overlap. Correcting for it does not
refine the published reading of five transcription factors; it
re-partitions them, moving a factor reported as a resolution failure to
the top and two reported as failures to at or below chance. A second and
more basic check --- the model's output on a fully masked input ---
shows that for one factor the perturbation curve has almost no range to
move in. Neither check requires an experiment. We provide both, together
with the closed forms and an implementation, and recommend that they be
reported.

\begin{center}\rule{0.5\linewidth}{0.5pt}\end{center}

\subsection{8. Reproducibility notes}

\begin{itemize}
\tightlist
\item
  Masking token is \textbf{N}, matching the source study. Using a pad
  token changes the deletion curve.
\item
  Every factor was fine-tuned under three random seeds. Validation AUC
  varies little between seeds: the spread is 0.004 to 0.017 (largest for
  TBP), against a between-factor spread of 0.077 and adjusted-recovery
  differences of 0.6. Seed variation is therefore not a plausible source
  of the between-factor differences we report.
\item
  Adjusted recovery is considerably less stable across seeds than
  validation AUC. Table~\ref{tab:t8} gives the per-factor values. The
  spread ranges from 0.022 to 0.220 and does not follow motif length.
  This is below the between-factor range of 0.61 but well above the
  within-run bootstrap half-width, which is 0.019 for CTCF.
  Recovery-rate comparisons at that scale therefore require repeated
  runs; see Section 6.8.
\end{itemize}

\begin{table}[!t]
\caption{Validation AUC across three seeds (HyenaDNA).}
\label{tab:t7}
\centering
\setlength{\tabcolsep}{4pt}
\small
\begin{tabular}{lccccc}
\toprule
TF & seed 0 & seed 1 & seed 2 & mean & spread \\
\midrule
CTCF  & 0.9670 & 0.9667 & 0.9716 & 0.9684 & 0.005 \\
GATA1 & 0.9718 & 0.9750 & 0.9714 & 0.9727 & 0.004 \\
MAX   & 0.9443 & 0.9488 & 0.9468 & 0.9466 & 0.005 \\
SP1   & 0.9063 & 0.9176 & 0.9091 & 0.9110 & 0.011 \\
TBP   & 0.8876 & 0.9044 & 0.8956 & 0.8959 & 0.017 \\
\bottomrule
\end{tabular}
\end{table}

\begin{table*}[!t]
\caption{Adjusted recovery across independent fine-tuning runs, held-out
chromosome 9. CTCF and GATA1 were run with five seeds, the remaining
three factors with three. Motif length $L$ is given for reference; the
spread does not follow it. Seed 0 was fine-tuned in a separate run from
the later seeds for all factors and is retained here for completeness;
excluding it changes the spreads by less than 0.01.}
\label{tab:t8}
\centering
\setlength{\tabcolsep}{4pt}
\footnotesize
\begin{tabular}{lccccccccccc}
\toprule
TF & $L$ & $K$ & seed 0 & seed 1 & seed 2 & seed 3 & seed 4 & mean & SD & spread \\
\midrule
CTCF  & 19 & 5 & $+0.569$ & $+0.551$ & $+0.678$ & $+0.540$ & $+0.503$ & $+0.568$ & 0.066 & 0.175 \\
GATA1 & 11 & 5 & $+0.491$ & $+0.482$ & $+0.487$ & $+0.316$ & $+0.271$ & $+0.409$ & \textbf{0.107} & \textbf{0.220} \\
MAX   & 10 & 3 & $+0.326$ & $+0.276$ & $+0.276$ & --- & --- & $+0.293$ & 0.029 & 0.050 \\
SP1   & 15 & 3 & $-0.006$ & $-0.022$ & $+0.001$ & --- & --- & $-0.009$ & 0.011 & 0.022 \\
TBP   & 15 & 3 & $-0.018$ & $-0.022$ & $+0.006$ & --- & --- & $-0.011$ & 0.015 & 0.028 \\
\bottomrule
\end{tabular}
\end{table*}

\begin{itemize}
\tightlist
\item
  The GC matching rule is a histogram match in bins of 0.02; 6 of 268
  factors failed to fill.
\item
  Code: \texttt{github.com/puckradi/gauge}, archived at
  \texttt{10.5281/zenodo.21594374}.
\end{itemize}

\begin{center}\rule{0.5\linewidth}{0.5pt}\end{center}

\FloatBarrier   


\begin{IEEEbiographynophoto}{Hyunkyung Han}
is a Ph.D. candidate in the Department of Integrative Medicine, Yonsei
University College of Medicine, Seoul, Republic of Korea. Her research concerns
interpretable and verifiable machine learning for medical imaging and genomic
sequence models, with an emphasis on the validity of evaluation procedures.
\end{IEEEbiographynophoto}

\begin{IEEEbiographynophoto}{Min Jung Kim}
is a Professor in the Department of Radiology and the Research Institute of
Radiologic Science, Yonsei University College of Medicine, Seoul, Republic of
Korea, and in the Department of Integrative Medicine. Her research interests
include breast imaging and the clinical evaluation of machine learning
systems.
\end{IEEEbiographynophoto}


\begin{thebibliography}{10}
\providecommand{\url}[1]{#1}
\csname url@samestyle\endcsname
\providecommand{\newblock}{\relax}
\providecommand{\bibinfo}[2]{#2}
\providecommand{\BIBentrySTDinterwordspacing}{\spaceskip=0pt\relax}
\providecommand{\BIBentryALTinterwordstretchfactor}{4}
\providecommand{\BIBentryALTinterwordspacing}{\spaceskip=\fontdimen2\font plus
\BIBentryALTinterwordstretchfactor\fontdimen3\font minus
  \fontdimen4\font\relax}
\providecommand{\BIBforeignlanguage}[2]{{%
\expandafter\ifx\csname l@#1\endcsname\relax
\typeout{** WARNING: IEEEtran.bst: No hyphenation pattern has been}%
\typeout{** loaded for the language `#1'. Using the pattern for}%
\typeout{** the default language instead.}%
\else
\language=\csname l@#1\endcsname
\fi
#2}}
\providecommand{\BIBdecl}{\relax}
\BIBdecl

\bibitem{zhou2026position}
S.~Zhou, M.~Huang, and K.~Li, ``Position: Genomic model research must move
  beyond anecdotal evaluation of interpretability methods,'' in \emph{Proc.
  43rd Int. Conf. Machine Learning (ICML)}, ser. Proceedings of Machine
  Learning Research, vol. 306, 2026, arXiv:2606.07607; code and data:
  \url{https://github.com/COLA-Laboratory/EvalXAI}.

\bibitem{kurth2026dnabert2}
I.~Kurth, P.~Yanez~Sarmiento, and B.~Y. Renard, ``Evaluating post-hoc
  explanations of the transformer-based genome language model {DNABERT-2},''
  \emph{arXiv preprint arXiv:2604.21690}, 2026.

\bibitem{hubert1985}
L.~Hubert and P.~Arabie, ``Comparing partitions,'' \emph{Journal of
  Classification}, vol.~2, no.~1, pp. 193--218, 1985.

\bibitem{altschul1985}
S.~F. Altschul and B.~W. Erickson, ``Significance of nucleotide sequence
  alignments: A method for random sequence permutation that preserves
  dinucleotide and codon usage,'' \emph{Molecular Biology and Evolution},
  vol.~2, no.~6, pp. 526--538, 1985.

\bibitem{jiang2008}
M.~Jiang, J.~Anderson, J.~Gillespie, and M.~Mayne, ``{uShuffle}: A useful tool
  for shuffling biological sequences while preserving the k-let counts,''
  \emph{BMC Bioinformatics}, vol.~9, p. 192, 2008.

\bibitem{bailey2011}
T.~L. Bailey, ``{DREME}: Motif discovery in transcription factor {ChIP-seq}
  data,'' \emph{Bioinformatics}, vol.~27, no.~12, pp. 1653--1659, 2011.

\bibitem{ferkingstad2015}
E.~Ferkingstad, L.~Holden, and G.~K. Sandve, ``Monte carlo null models for
  genomic data,'' \emph{Statistical Science}, vol.~30, no.~1, pp. 59--71, 2015.

\bibitem{domanska2018}
D.~Domanska, C.~Kanduri, B.~Simovski, and G.~K. Sandve, ``Mind the gaps:
  Overlooking inaccessible regions confounds statistical testing in genome
  analysis,'' \emph{BMC Bioinformatics}, vol.~19, no.~1, p. 481, 2018.

\bibitem{kanduri2019}
C.~Kanduri, C.~Bock, S.~Gundersen, E.~Hovig, and G.~K. Sandve, ``Colocalization
  analyses of genomic elements: Approaches, recommendations and challenges,''
  \emph{Bioinformatics}, vol.~35, no.~9, pp. 1615--1624, 2019.

\bibitem{salvatore2020}
S.~Salvatore, K.~D. Rand, I.~Grytten, E.~Ferkingstad, D.~Domanska, L.~Holden,
  M.~Gheorghe, A.~Mathelier, I.~Glad, and G.~K. Sandve, ``Beware the jaccard:
  The choice of similarity measure is important and non-trivial in genomic
  colocalisation analysis,'' \emph{Briefings in Bioinformatics}, vol.~21,
  no.~5, pp. 1523--1530, 2020.

\bibitem{kanduri2025}
C.~Kanduri, M.~Mamica, E.~W. Olstad, M.~Zucknick, J.~J. Li, and G.~K. Sandve,
  ``Beware of counter-intuitive levels of false discoveries in datasets with
  strong intra-correlations,'' \emph{Genome Biology}, vol.~26, no.~1, p. 249,
  2025.

\bibitem{de2014}
S.~De, B.~S. Pedersen, and K.~Kechris, ``The dilemma of choosing the ideal
  permutation strategy while estimating statistical significance of genome-wide
  enrichment,'' \emph{Briefings in Bioinformatics}, vol.~15, no.~6, pp.
  919--928, 2014.

\bibitem{chikina2012}
M.~D. Chikina and O.~G. Troyanskaya, ``An effective statistical evaluation of
  {ChIPseq} dataset similarity,'' \emph{Bioinformatics}, vol.~28, no.~5, pp.
  607--613, 2012.

\bibitem{proovErlap}
N.~Gualandi, A.~Bertozzo, and C.~Brancolini, ``{ProOvErlap}: Assessing feature
  proximity/overlap and testing statistical significance from genomic
  intervals,'' \emph{Journal of Biological Chemistry}, vol. 301, no.~6, p.
  110209, 2025.

\bibitem{hanhijarvi}
S.~Hanhij{\"a}rvi, G.~C. Garriga, and K.~Puolam{\"a}ki, ``Randomization
  techniques for graphs,'' in \emph{Proc. SIAM Int. Conf. Data Mining (SDM)},
  2009.

\bibitem{koo2019}
P.~K. Koo and S.~R. Eddy, ``Representation learning of genomic sequence motifs
  with convolutional neural networks,'' \emph{PLoS Computational Biology},
  vol.~15, no.~12, p. e1007560, 2019.

\bibitem{hyenadna}
E.~Nguyen, M.~Poli, M.~Faizi, A.~Thomas, M.~Wornow, C.~Birch-Sykes,
  S.~Massaroli, A.~Patel, C.~Rabideau, Y.~Bengio, S.~Ermon, C.~R{\'e}, and
  S.~Baccus, ``{HyenaDNA}: Long-range genomic sequence modeling at single
  nucleotide resolution,'' in \emph{Advances in Neural Information Processing
  Systems (NeurIPS)}, vol.~36, 2023.

\bibitem{dnabert2}
Z.~Zhou, Y.~Ji, W.~Li, P.~Dutta, R.~Davuluri, and H.~Liu, ``{DNABERT-2}:
  Efficient foundation model and benchmark for multi-species genomes,'' in
  \emph{Proc. Int. Conf. Learning Representations (ICLR)}, 2024.

\bibitem{integratedgradients}
M.~Sundararajan, A.~Taly, and Q.~Yan, ``Axiomatic attribution for deep
  networks,'' in \emph{Proc. 34th Int. Conf. Machine Learning (ICML)}, ser.
  Proceedings of Machine Learning Research, vol.~70, 2017, pp. 3319--3328,
  arXiv:1703.01365.

\bibitem{darteval}
A.~Patel, A.~Singhal, A.~Wang, A.~Pampari, M.~Kasowski, and A.~Kundaje,
  ``{DART-Eval}: A comprehensive {DNA} language model evaluation benchmark on
  regulatory {DNA},'' in \emph{Advances in Neural Information Processing
  Systems (NeurIPS), Datasets and Benchmarks Track}, 2024, arXiv:2412.05430.

\end{thebibliography}
\end{document}